\documentclass[prl,twocolumn,showpacs,amsfonts,superscriptaddress,
amsmath,floatfix]{revtex4}
\usepackage{graphicx}
\usepackage{psfrag}
\newcommand{\Journal}[4]{#1 {\bf #2}, #3 (#4)}
\newcommand{\PR}{Phys. Rev.}
\newcommand{\PRL}{Phys. Rev. Lett.}
\newcommand{\PRA}{Phys. Rev. A}

\newcommand{\JMP}{J. Math. Phys.}

\newcommand{\Science}{Science}
\newcommand{\PLA}{Phys. Lett. A}
\begin{document}
\title {Rotating ground states of a one-dimensional spin-polarized gas of
fermionic atoms with attractive p-wave interactions on a mesoscopic ring}
\author{M. D. Girardeau}
\email{girardeau@optics.arizona.edu}
\affiliation{College of Optical Sciences, University of Arizona,
Tucson, AZ 85721, USA}
\author{E. M. Wright}
\email{ewan.wright@optics.arizona.edu}
\affiliation{College of Optical Sciences, University of Arizona,
Tucson, AZ 85721, USA}
\affiliation{Department of Physics, University of Arizona,
Tucson, AZ 85721, USA}
\date{\today}
\begin{abstract}
The major finding of this paper is that a one-dimensional spin-polarized gas comprised of an even number of
fermionic atoms interacting via attractive p-wave interactions and confined to a mesoscopic ring has a degenerate pair of ground states that are oppositely rotating.  In any realization the gas will thus spontaneously rotate one way or the other in spite of the fact that there is no external rotation or bias fields.  Our goal is to show that this counter-intuitive finding is a natural consequence of the combined effects of quantum statistics, ring topology, and exchange interactions.
\end{abstract}
\pacs{03.75.-b,05.30.Fk}
\maketitle
The rapidly increasing sophistication of experimental techniques for probing
ultracold gases has caused a shift of emphasis in theoretical and experimental
work in recent years, from effective field approaches to more refined
methods capable of dealing with strong correlations. Such strong correlations arise, for example, in ultracold gases
confined in de Broglie waveguides with transverse trapping
so tight that the atomic dynamics is essentially one-dimensional (1D) \cite{Ols98},
with confinement-induced resonances \cite{Ols98,GraBlu04}
allowing Feshbach resonance tuning \cite{Rob01} of the effective 1D
interactions to very large values. This has allowed for the experimental verification
\cite{Par04Kin04,Kin05,Kin06} of the fermionization of bosonic ultracold
vapors in such geometries predicted by the Fermi-Bose (FB)
mapping method \cite{Gir60Gir65}.
The ``fermionic Tonks-Girardeau'' (FTG) gas \cite{GirOls03,GirNguOls04}, a
spin-aligned Fermi gas with very strong \emph{attractive} 1D odd-wave
interactions, can be realized by a 3D p-wave Feshbach resonance as, e.g., in
ultracold $^{40}$K vapor \cite{Tik04}. It has been pointed out
\cite{GraBlu04,GirOls03,GirNguOls04} that the generalized FB mapping
\cite{CheShi98,GraBlu04,GirOls03,GirNguOls04} can be exploited in the opposite
direction to map the \emph{ideal FTG} gas with infinitely strong attractive p-wave interactions to the \emph{ideal Bose} gas,
leading to ``bosonization'' of many properties of this Fermi system.
One of us (MDG) in collaboration with A. Minguzzi showed recently \cite{GirMin06} that for an even number $N$ of fermions on a mesoscopic ring,
the ideal FTG ground state is highly degenerate due to fragmentation
of the mapped ideal Bose gas between
two macroscopically occupied orbitals with circumferential wave vectors
$\pm\pi/L$, where $L$ is the ring circumference.

The purpose of the present paper is to examine how the ground state properties of a FTG gas on a mesoscopic ring are changed for an even number $N$ of fermions when the strength of the atom-atom attraction is made finite, although still very large, in which case there exists an FB mapping between the FTG gas and a system of $N$ bosons with repulsive delta-function interactions. Based on this we have found the remarkable result that whilst most of the ground state degeneracy of the ideal FTG gas is lifted a twofold degeneracy remains, and these two states have \emph{nonzero total angular momentum} $\pm\frac{1}{2} N\hbar$, corresponding to BEC of all $N$ particles of the mapped Bose gas into either an orbital of angular momentum $\frac{\hbar}{2}$ or one of angular momentum $-\frac{\hbar}{2}$.  The lifting of the degeneracy will be given a natural interpretation in terms of the mapped Bose system using the famous no-fragmentation theorem of Nozieres and Saint James \cite{NSJ82}. Thus, in any given experimental realization with an even number $N$ of fermions the FTG will spontaneously rotate one way or the other around the ring with equal probability. This apparently violates a theorem of F. Bloch \cite{Boh49} requiring the ground state to be non-rotating, and we shall demonstrate why the theorem fails for our case of fermions on a mesoscopic ring.

{\it Ideal FTG gas:} To set the stage we consider the ideal FTG gas in general and on a ring. The ideal FTG gas consists of
spin-aligned fermions with infinitely-strong zero-range attractions which are
a zero-width,
infinite depth limit of a square well of depth $V_0$ and width $2x_0$, with
the limit taken such that $V_0 x_0^2\to(\pi\hbar^2)/8m_{eff}$ where $m_{eff}$ is the
effective mass of the colliding pair \cite{GraBlu04,GirOls03,GirNguOls04}.
It causes odd-wave scattering (1D analog of 3D p-wave scattering) with 1D
scattering length $a_{1D}=-\infty$, with the result that all energy eigenstates
$\Psi_F$ of the FTG gas are obtained from corresponding ideal Bose gas states
$\Psi_B$ by the mapping $\Psi_B\to\Psi_F=A\Psi_B$, where
$A$ is the same mapping function used in the original
solution of the TG gas \cite{Gir60Gir65}:
$A(x_1,\cdots,x_N)=\prod_{1\le j<\ell\le N}\text{sgn}(x_j-x_\ell)$
where the sign function $\text{sgn}(x)$ is $+1\ (-1)$ if $x>0\ (x<0)$. This
introduces sign-changing discontinuities in the FTG states necessary to
reconcile fermionic antisymmetry with a strong interaction
in the zero-range limit $x_0\to 0$ \cite{CheShi98}. Inside the square well the
solution passes smoothly through a zero at $x_j-x_\ell=0$, so the discontinuity
is an illusion of the zero-range limit
\cite{CheShi98,GraBlu04,GirOls03,GirNguOls04}. The exact ground state is
\begin{equation}\label{FTG ground}
\Psi_{F0}(x_{1},\cdots,x_{N})=
A(x_{1},\cdots,x_{N})\prod_{j=1}^{N}\phi_{0}(x_{j}),
\end{equation}
where $\phi_{0}(x)$ is the lowest single-particle orbital for the given
boundary conditions, illustrating the mapping from the ideal Bose gas
ground state $\Psi_{B0}=\prod_{j=1}^{N}\phi_{0}(x_{j})$ to the fermionic
FTG ground state $\Psi_{F0}$. The physical meaning of this mapping is clarified
by looking at the solution of the two-fermion problem both inside
and outside the square well, before passing to the limit $x_0\to 0$.
Denoting the relative coordinate by $x_{12}=x_1-x_2$, the exterior solution
($|x_{12}|>x_0$) in the ideal FTG limit (1D scattering length
$a_{1D}\to -\infty$)
represents a zero-energy scattering resonance, with wave function +1 for
$x_{12}>0$ and -1 for $x_{12}<0$, and the interior solution ($|x_{12}|<x_0$)
fitting smoothly onto this is $\sin(\kappa x_{12})$ with
$\kappa=\sqrt{mV_{0}/\hbar^2}=\pi/2x_0$. The corresponding mapped Bose gas
state is +1 everywhere outside the well (ideal Bose gas ground state),
but inside the well it is $\sin(\kappa |x_{12}|)$.
Since this vanishes with a cusp at
$x_{12}=0$, physical consistency requires the presence of a zero-diameter
hard core interaction added to the square well. The mapped Bose
gas is then not truly ideal, but rather a \emph{TG gas with
superimposed attractive well}, whose nontrivial interior wave function becomes
invisible in the zero-range limit, simulating an ideal Bose gas
insofar as the energy and exterior wave function are concerned. The required
impenetrable core is physically quite reasonable, since the atoms have a strong
short-range Pauli exclusion repulsion of their inner shells, whose diameter is
effectively zero and strength infinite on length and energy scales appropriate
to ultracold gas experiments.

{\it Ideal FTG gas on a ring:} Here we review the discussion of Girardeau and
Minguzzi for fermions on a ring of circumference $L$ \cite{GirMin06}. For an
odd number $N$ of fermions Eq. (\ref{FTG ground}) is the correct FTG ground
state, with $\phi_0=1/\sqrt{L}$ the trivial and unique ground state orbital, a
plane-wave with zero wave vector $k$ and hence zero angular momentum.
However, if $N$ is even we encounter a difficulty. For
example, for $N=2$ and $0<x_2<L$
the mapping function $A=\text{sgn}(x_1-x_2)$ is -1 for
$x_1=0+\epsilon$ but +1 for $x_1=L-\epsilon$, where $\epsilon\to 0+$.
As a result, $\Psi_{F0}$ is not periodic but instead antiperiodic,
violating the requirement that the fermion wave function be single valued.
More generally, $A(x_{1},\cdots,x_{N})$ is antiperiodic for all even $N$,
leading to the same difficulty. This problem is easily repaired by taking
the Bose ground state $\Psi_{B0}$ to be the lowest \emph{anti}periodic
state, a \emph{fragmented} BEC with $wN$ atoms in the orbital $e^{i\pi x_j/L}$
and $(1-w)N$ in $e^{-i\pi x_j/L}$ with $0\le w\le 1$. The fermionic ground
state $\Psi_{F0}=A\Psi_{B0}$ is then properly periodic, but is $(N+1)$-fold
degenerate and conveniently labeled by a quantum number
$\ell_z=(w-\frac{1}{2})N=0,\pm 1,\pm 2,\cdots,\pm \frac{N}{2}$
related to the eigenvalue $L_z$ of orbital angular momentum by
$L_z=\ell_z\hbar$. Since $N$ is even, $L_z$ takes on all \emph{integral} values
from $-\frac{1}{2} N\hbar$ to $\frac{1}{2} N\hbar$, in spite of the
half-integral angular momenta of the two orbitals, and the fermionic
ground states have the same angular momentum spectrum since the total angular
momentum operator commutes with the mapping function $A$.

The above discussion leading to rotating and degenerate ground states clearly relies on the mesoscopic nature of the ring as it relies on the discreteness of the angular momentum.  In contrast, Bloch's ``no-rotation'' theorem \cite{Boh49} is based on a Galilean transformation treating the momentum as a continuous variable and therefore holds strictly only in the thermodynamic limit; that is why our mesoscopic system can violate Bloch's theorem.  Furthermore, the physics underlying why FTG gases with an even number of fermions on a ring can have rotating ground states is also revealed in the above discussion: It is a consequence of the combined effects of quantum statistics, which is implicit in the use of the mapping function $A(x_{1},\cdots,x_{N})$ to incorporate the anti-symmetry of the fermion wave function, and the ring topology which necessitates a non-zero angular momentum so that the underlying Bose ground state can exhibit the antiperiodic boundary conditions needed to counter the antiperiodicity of $A$.  The same general argument applies also for the case of finite interactions since the problem can still be approached using the same mapping function $A$ (see below), albeit that the mapped Bose problem is no longer ideal. However, for the ideal FTG it is still possible to consider combinations of the degenerate ground states that have zero angular momentum and to address this we need to consider the effects of finite interactions.

\begin{figure}[t]
  \centering
\includegraphics[width=7.5cm,angle=0]{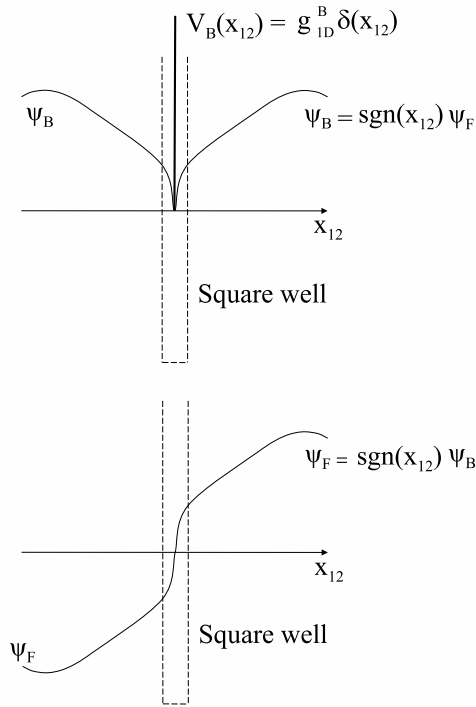}
  \caption{Two-particle fermionic wave function $\Psi_F$ and mapped
bosonic wave function $\Psi_B$ as a function of $x_1$ for fixed $x_2$.
The potential for both $\Psi_F$ and $\Psi_B$ is a deep and narrow square
well plus a point hard core, but in the zero-range limit its effect on $\Psi_B$
outside the well is the same as that of $v_B=g_{1D}^B\delta(x_{12})$.}
  \label{fig1}
\end{figure}

{\it FTG with finite interactions:} It has been realized in recent
years that the FB mapping method used to exactly solve the TG gas
\cite{Gir60Gir65} and FTG gas \cite{GirOls03,GirNguOls04} is not restricted
to the TG case of point hard core boson-boson repulsion and the FTG case of
infinite zero-range fermion-fermion attraction. In fact, exactly the same
unit antisymmetric mapping function $A(x_1,\cdots,x_N)$ used there also
provides an exact mapping between bosons with delta-function repulsions
$g_{1D}^B\delta(x_j-x_\ell)$ of any strength, i.e., the Lieb-Liniger (LL)
model \cite{LL63}, and spin-aligned fermions with attractive interactions
of a generalized FTG form with reciprocal fermionic coupling constant
$g_{1D}^F$.  The 1D scattering
length $a_{1D}<0$ is invariant under the mapping, which, therefore,
maps strongly attractive fermions to weakly repulsive bosons and
strongly repulsive bosons to weakly attractive fermions, the coupling
constants being related to the scattering length and each other by
$g_{1D}^B=2\hbar^2/m|a_{1D}|$, $g_{1D}^F=2\hbar^2|a_{1D}|/m$, and
$g_{1D}^Bg_{1D}^F=\frac{4\hbar^4}{m^2}$
\cite{GraBlu04,GirOls03,GirNguOls04,CheShi98}. The physical meaning of this
mapping is illustrated by Fig.\ref{fig1}, which shows, for the case $N=2$,
how the mapping converts a FTG attractive interaction
to the LL delta function interaction, which generates cusps in the wave
function in accord with the LL contact conditions \cite{LL63}. The potential,
both for $\Psi_F$ and $\Psi_B$, consists of a deep and narrow square well of
width $2x_0$ and depth $V_0$ plus a TG impenetrable core at $x_{12}=0$,
as discussed previously for the ideal FTG limit $a_{1D}\to -\infty$.
The point core has no effect on
$\Psi_F$, which already vanishes at $x_{12}=0$ due to antisymmetry, but
it is necessary for $\Psi_B$ for consistency of the mapping. The generalized
FTG well involves going to the zero width limit $x_0\to 0$ and infinite
depth limit $V_0\to\infty$ with $V_0x_0^2$ held constant as for the FTG
limit $a_{1D}\to-\infty$, but now at a slightly smaller value, such that
the parameter $\kappa$ in the interior wave function $\sin(\kappa x_{12})$
scales as $\kappa=\frac{\pi}{2x_0}-\frac{2}{\pi|a_{1D}|}$ as
$x_0\to 0$ \cite{GirOls03,GirNguOls04}. Outside the well this potential
has the same effect on $\Psi_B$ as a LL delta function
$g_{1D}^B\delta(x_{12})$; this is the physical meaning of the mapping from
the spin-aligned Fermi gas to the LL Bose gas.
Here we will use this mapping to treat the case of spin-aligned fermions
with strong but finite attraction via mapping to an LL gas with weak
repulsions.

The relative strengths of the FTG attractive p-wave interactions and the LL
repulsive interactions are quantified in the dimensionless coupling
coefficients $\gamma_F=\frac{mg_{1D}^Fn}{\hbar^2}$ and
$\gamma_B=\frac{mg_{1D}^B}{\hbar^2 n}$, respectively, where $n=\frac{N}{L}$ is
the linear atomic density, and $\gamma_F\gamma_B=4$.  The ideal FTG gas is
realized in the limit $a_{1D}\to-\infty$ so that $\gamma_F\to\infty$ and
$\gamma_B\to 0$.  Here we are interested in the limit of finite but still very
strong p-wave attractive interactions, or the limit $\gamma_F>>1$ and
$\gamma_B<<1$.  Moreover, this is precisely the limit where the ground state
properties of the LL Hamiltonian can be accurately captured using mean field
theory in which all N bosons are assumed to occupy the same single particle
orbital $\phi$ that is determined self-consistently via the solution of the
Gross-Pitaevskii (GP) equation \cite{LifPit89}.
The $N$-particle boson wave function is then written in second quantized form as $|\Psi_{B}\rangle=(\hat{a}^\dagger)^{N}|0\rangle/\sqrt{N!}$, with $\hat{a}=\int dx\phi^*(x)\hat{\psi}(x)$, $\hat{\psi}^\dagger(x)$ and $\hat{\psi}(x)$ being boson creation and annihilation field operators.  Lieb {\it et al.} \cite{LieSeiYng03} have studied the parameter space in which a system of $N$ bosons with repulsive interactions in a 3D trap with tight transverse binding may be treated as an effective 1D LL gas, and the GP approach should be valid in the limit $\gamma_B<1/N^2$, the same criterion applying to our ring trap with tight transverse binding.

In order to address fragmentation of the ground state of the FTG gas in the
presence of finite p-wave interactions, here we adopt a more general
Hartee-Fock approach to the mapped LL problem in the limit $\gamma_B\ll 1$,
which is a two-component generalization of the GP theory, where the two
components are the two degenerate ground state orbitals
$\phi_{\pm}(x)=e^{\pm i\pi x/L}/\sqrt{L}$ of the GP equation
with antiperiodic boundary conditions and energies
$E_\pm=\frac{(\pi\hbar)^2}{2mL^2}N+\frac{g}{L}N(N-1)$ where
$g\equiv g_{1D}^B$. The exact second-quantized LL Hamiltonian is
\begin{equation}\label{Exact H}
\hat{H}=\int_0^{L}dx\left\{-\frac{\hbar^2}{2m}\hat{\psi}^\dagger(x)
\frac{\partial^2}{\partial x^2}\hat{\psi}(x)
+\frac{g}{2}[\hat{\psi}^\dagger(x)]^2[\hat{\psi}(x)]^2\right\},
\end{equation}
and we evaluate its expectation value in the $(N+1)$ states
\begin{equation}\label{states}
|\Psi_{B}(w)\rangle=
\frac{(\hat{a}_+^\dagger)^{wN}(\hat{a}_-^\dagger)^{(1-w)N}}
{\sqrt{(wN)![(1-w)N]!}}|0\rangle,
\end{equation}
where $\hat{a}_\pm=\int dx\phi_\pm^*(x)\hat{\psi}(x)$. The states in Eq. (\ref{states}) represent fragmented states of the two orbitals $\phi_\pm(x)$ in second-quantized form, and are the analogue of the previously discussed fragmented states $\Psi_{B0}$ associated with the ideal FTG gas.
The energy expectation values of the LL Hamiltonian (\ref{Exact H}) taken with
respect to the fragmented states (\ref{states}) are given by
\begin{equation}\label{MF}
E_0(w)=\frac{(\pi\hbar)^2}{2mL^2}N+\frac{g}{2L}[-2N^2w^2+2N^2w+N^2-N]\ .
\end{equation}
The energy assumes a maximum at $w=\frac{1}{2}$, representing a fragmented 
state with zero total angular momentum and equal
occupations $\frac{N}{2}$ of the orbitals $\phi_\pm$,
and the energy minima are at $w=1$ and $w=0$, the unfragmented states with
all $N$ atoms in either $\phi_+$ or $\phi_-$, energies
$E_0(1)=E_0(0)=\frac{(\pi\hbar)^2}{2mL^2}N+\frac{g}{2L}N(N-1)$,
and total angular momenta $\pm\frac{N\hbar}{2}$. These two rotating states lie
lower than the nonrotating state $w=\frac{1}{2}$ by an amount
$\frac{g}{4L}N^2$, the extra exchange energy arising from fragmentation of the
state $w=\frac{1}{2}$, in accordance with the no-fragmentation theorem
\cite{NSJ82}, which can be regarded as a physical mechanism forcing the ground LL state to be rotating.  Since the total energy and total angular momentum operators commute with the mapping function $A$, it follows that the FTG gas has two degenerate ground states with opposite angular momenta $\pm\frac{N\hbar}{2}$, and this is the main result of this paper.

It is instructive to compare these mean field results with exact solutions of
the LL Bethe ansatz equations \cite{LL63} for $N=2$ and antiperiodic boundary
conditions. These states have the form
$\Psi(x_1,x_2)=e^{ik_1x_1}e^{ik_2x_2}-e^{i\theta}e^{ik_2x_1}e^{ik_1x_2}$
in the sector $0<x_1<x_2<L$, and are equal to $\Psi(x_2,x_1)$ when
$0<x_2<x_1<L$. For $\gamma_B\ll 1$ the lowest nonrotating state has
$k_1=-k$ and $k_2=k$ where $kL=\pi+\epsilon$ with $0<\epsilon\ll 1$, and
we find $\epsilon=\frac{2\gamma_B}{\pi}+\mathit{O}(\gamma_B^2)$. Its
energy is $2\left(\frac{\pi}{L}\right)^2+\frac{8\gamma_B}{L^2}
+\mathit{O}(\gamma_B^2)$, which agrees to order $\frac{\gamma_B}{L^2}$ with
the mean field energy
$E_0(\frac{1}{2})$ of Eq. (\ref{MF}), with LL units $\hbar=2m=1$ and
$g=2c=\frac{4\gamma_B}{L}$. The twofold degenerate ground states
$\Psi_0^{\pm}$ have $k_1L=\pm\pi-\epsilon$ and $k_2L=\pm\pi+\epsilon$
with $0<\epsilon\ll 1$, and we find $\epsilon=\sqrt{2\gamma_B}
+\mathit{O}(\gamma_B)$ and energy
$2\left(\frac{\pi}{L}\right)^2+\frac{4\gamma_B}{L^2}$, which again
agrees with the mean field result $E_0(0)=E_0(1)$ of Eq. (\ref{MF})
to order $\frac{\gamma_B}{L^2}$.
We conclude that our mean field solutions capture both the qualitative and
quantitative features of the exact antiperiodic LL solutions for
$\gamma_B\ll 1$, and in particular, the exact LL ground state is twofold
degenerate and rotating, with angular momenta $\pm\frac{N\hbar}{2}$.

A potential scheme to measure the angular momentum of the rotating gas is
the non-destructive approach discussed in Ref. \cite{MarZhaWri97}, which 
employs Raman transitions between hyperfine levels of the atoms due to 
oppositely circularly polarized laser fields that propagate along a direction 
in the plane of the atomic ring. The spatial absorption image for the circular 
polarizations can depend on the angular momentum state of the gas since the 
rotation gives rise to a Doppler shift that modifies the probability of the 
Raman transition.  Then if the gas is rotating the absorption image will show 
an asymmetry with respect to the axis passing through the center of the ring, 
since the atoms on either side of the axis will be moving in opposite 
directions, experience different Doppler shifts, and hence produce different 
absorptions.

{\it Ring dark solitons:} So far we have tacitly assumed that the density profile of the ground state of the FTG gas, and the underlying LL gas, is spatially homogeneous.  Relaxing this assumption raises the possibility that a non-rotating ground state may be realized in the form of a ring dark soliton.  Dark solitons are real valued solutions $\phi_{DS}(x)$ of the GP equation
\cite{CarClaRei00}, neglecting any overall phase factor, which have an inhomogeneous density profile and a phase change of $\pi$ at the position where the density touches zero. Thus dark solitons have the required antiperiodicity for the mapped Bose solutions. Details of how to calculate the ring dark solitons for finite interactions are given exhaustively in Ref. \cite{CarClaRei00}, and for the ideal FTG the dark soliton solution of the GP equation
with $(g=0)$ degenerates into $\phi_{DS}(x)=\sqrt{\frac{2}{L}}\sin(\pi x/L)$, and is degenerate with the previous rotating plane-wave solutions. When the ring dark solitons $(g>0)$ arise they always have energies higher than the rotating plane-wave solutions, so that although they are non-rotating they represent excited states.  An intuitive way to convey this is that the healing length associated with the ring dark solitons always engenders a larger kinetic energy penalty than the wavevectors $\pm\pi/L$ associated with the rotating ground states.

In summary, we predict that the ground state of an FTG gas with an even number of fermions on a ring and finite p-wave interactions is doubly degenerate and rotating.  We have further shown that the rotation is a macroscopic manifestation of the strong microscopic many-body correlations present in the gas.  Observation of the spontaneous rotation of an FTG gas on a ring would constitute a striking validation of the underlying strong correlations.
%
%

%
\end{document}